\documentclass[reprint,prb,aps,amsmath,amssymb,floatfix]{revtex4-1}
\usepackage{graphicx}
\usepackage{epsfig}

\begin{document}
\title{Heat current across a capacitively coupled double quantum dot}

\author{A. A. Aligia}
\affiliation{Centro At\'{o}mico Bariloche, Comisi\'{o}n Nacional
de Energ\'{\i}a At\'{o}mica, 8400 Bariloche, Argentina}
\affiliation{Instituto Balseiro, Comisi\'{o}n Nacional
de Energ\'{\i}a At\'{o}mica, 8400 Bariloche, Argentina}
\affiliation{Consejo Nacional de Investigaciones Cient\'{\i}ficas y T\'ecnicas,
1025 CABA, Argentina}
\email{aligia@cab.cnea.gov.ar}

\author{D. P\'erez Daroca}
\affiliation{Gerencia de Investigaci\'on y Aplicaciones, Comisi\'on Nacional de
Energ\'ia At\'omica, 1650 San Mart\'{\i}n, Buenos Aires, Argentina}
\affiliation{Consejo Nacional de Investigaciones Cient\'{\i}ficas y T\'ecnicas,
1025 CABA, Argentina}

\author{Liliana Arrachea}
\affiliation{International Center for Advanced Studies, Escuela de Ciencia y Tecnolog\'{\i}a, 
Universidad Nacional de San Mart\'{\i}n,  25 de Mayo y Francia, 1650 Buenos Aires, Argentina}
\affiliation{Consejo Nacional de Investigaciones Cient\'{\i}ficas y T\'ecnicas,
1025 CABA, Argentina}

\author{P. Roura-Bas}
\affiliation{Centro At\'{o}mico Bariloche, Comisi\'{o}n Nacional
de Energ\'{\i}a At\'{o}mica, 8400 Bariloche, Argentina}
\affiliation{Consejo Nacional de Investigaciones Cient\'{\i}ficas y T\'ecnicas,
1025 CABA, Argentina}

\begin{abstract}
We study the heat current through two capacitively coupled quantum dots  
coupled in series with two conducting leads at different  temperatures $T_L$ and $T_R$
in the spinless case (valid for a high applied magnetic field). 
Our results are also valid for the heat current 
through a single quantum dot with strongly ferromagnetic leads pointing in opposite directions 
(so that the electrons with given spin at the dot can jump only to one lead) or through a quantum dot 
with two degenerate levels with destructive quantum interference 
and high magnetic field. 
Although the charge current is always zero, the heat current is finite when the interdot Coulomb repulsion $U$ is  taken into account due to  many-body effects. 
We study the thermal conductance as a function of  temperature and 
the dependence of the thermal current with the couplings to the leads, $T_L-T_R$, energy levels of
the dots and $U$, including conditions for which an orbital Kondo regime takes place.  
When the energy levels of the dots are different, the device has rectifying properties 
for the thermal current. We find that the ratio between the thermal current resulting from a thermal bias $T_L>T_R$ and the one from $T_L<T_R$ is maximized for particular values of the energy levels, 
one above and the other below the Fermi level.
\end{abstract}


\maketitle

\section{Introduction}

In the last years, there has been a significant interest in the thermal transport in the quantum coherent regime of electron systems. 
A prominent example is the experimental measurement of the quantum of thermal conductance in the quantum Hall regime. \cite{jezouin}
In the most common scenario the thermal transport comes along with charge transport, which is at the heart of the basic Wiedemann-Franz law. The interplay between electrical and thermal transport is also the key of thermoelectricity, which is a very active avenue of research. 
\cite{beni,engine,miao,svila,sierra,craven,dutta,cui,kra,pola,costi,leij,linke,anderg,aze,corna,guo,tooski,ng,aze2,kim,asha,rincon,dorda,diego,ecke,erdman,klo,sierra1,dare1,jiang,cui2,bala,rouarr,li,erdman2,karki,ridley,dare2}

\begin{figure}[tbp]
\includegraphics[clip,width=\columnwidth]{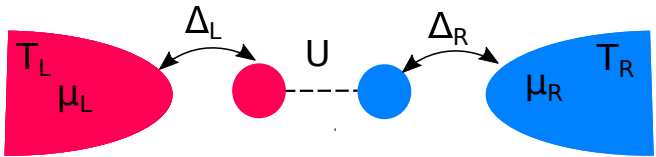}
\caption{(Color online) Sketch of the system analyzed in this work in which two capacitively quantum dots are attached 
to two conducting leads of  spinless electrons and at different temperatures and chemical potentials. 
It also describes two additional models as explained in the main text.}
\label{esquema}
\end{figure}

Heat without charge transport is also related to interesting effects, including energy harvesting with quantum dots \cite{rafa1,rafa2,jaliel} and rectification.\cite{ruoko,yada} The relevant mechanism is the capacitive coupling of charge due to the Coulomb interaction. Interestingly, the quantum of thermal conductance per ballistic channel is 

\begin{equation}
\kappa_{0}= \pi^2k_B^2 T/(3h),  \label{quantum}
\end{equation}
and is independent of the statistics of the particles \cite{pendry}.  
In general, Eq. (\ref{quantum}) is an upper bound for the conductance.
Capacitive couplings in electron systems are expected to achieve
significantly lower values than this bound, since it is very unlikely to realize a perfect ballistic regime in this scenario. \cite{hugo} A paradigmatic device to analyze the behavior of the thermal transport mediated by the Coulomb interaction consists of two capacitively coupled quantum dots (QDs) in contact to reservoirs $L$ and $R$ at
different temperatures $T_L$ and $T_R$, as sketched in Fig. \ref{esquema}. The simplest situation corresponds to single-level QDs of spinless electrons, which could be realized in the presence of a magnetic field. The Coulomb interaction is denoted 
by $U$. Electrons do not tunnel between the two QDs. Hence, the thermal current is not accompanied by any charge current. A physical picture for the thermal transport in this device is given in Section \ref{pict}. See also Ref. \onlinecite{ruoko}. 

The model is equivalent to that of transport between two levels with destructive interference
under high magnetic fields.\cite{desint,note} 
It is also equivalent to a spinfull model for one dot in which electrons with spin up can only 
hop to the left lead and electrons with spin down can only hop to the right lead or vice versa. Such a configuration ca be realized for totally polarized ferromagnetic leads with opposite orientation
(angle $\pi$ between them).\cite{pola} 

The system was previously studied by recourse to rate equations, in the regime where the coupling between the QDs and the reservoirs is negligible, compared to $U$ as well as $k_B T_L$ and $k_B T_R$ \cite{rafa1,rafa2,ruoko}. More recently,
results were also presented for arbitrary coupling to the reservoirs and small $U$ and/or high $T$. \cite{yada} There is another interesting regime in this system, which consists in an orbital Kondo regime, taking place 
below the characteristic temperature $T_K $. 

The Kondo effect is one of the most paradigmatic phenomenon in strongly
correlated condensed matter systems.\cite{hewson-book} 
Its simplest version is realized in a single spinfull QD, which behaves as a quantum impurity when it is occupied by a single electron. It is characterized by the emergence of
a many-body singlet below $T_K$, which is formed by  the spin 1/2 localized at the impurity and the spin
1/2 of the conduction electrons near the Fermi level. 
As a consequence the spectral density 
of the impurity displays a resonance at the Fermi energy. This explains the widely observed 
zero-bias anomaly in charge transport through quantum dots with an odd number of 
electrons.\cite{svila,sierra,dutta,costi,gold,cro,wiel,liang} The Kondo effect with spin $S>1/2$ has also been observed.\cite{roch,parks,serge}  
The role of the impurity spin can be replaced by other quantum degree of
freedom that distinguishes degenerate states, such as orbital momentum. 
Orbital degeneracy leads to the orbital Kondo effect or to more exotic Kondo effects, like the SU(4) one, when both orbital and spin degeneracy coexist. Some examples are present in nanoscopic systems.\cite{aze,corna,jari,ander,buss,tetta,grove,mina,lobos,3ch} 
Evidence of the orbital Kondo effect has also been observed in magnetic systems in which the spin degeneracy is broken.\cite{kole,adhi,kov}

In the case of  the double-dot model  of Fig. \ref{esquema} the occupancy of
one dot or the other plays the role of the spin, 
and a many-body state develops below $T_K$. 
The study of the thermal transport in this regime  has not been
addressed so far and one of the aims of the present work is to cover this gap. Concretely, we will focus on the regime of high temperature and analyze the effect of finite coupling between QDs and the reservoirs, 
as well as the low temperature regime below $T_K$. In this regime, it is not possible to calculate the heat current exactly. For this reason, we rely on different approximations for systems out of equilibrium within the present state of the art techniques, which are described in Section \ref{scoup}.
Most of the results presented for finite coupling to the reservoirs
were obtained using non-equilibrium perturbation theory up to second order in $U$,
which is valid for small or moderate values of $U$.\cite{hersh,none}
For infinite $U$ we use renormalized perturbation theory,\cite{ng,hbo,cb,ct,ogu2} 
and the non-crossing approximation.\cite{win,roura_1,tosib}
We carefully analyze in each case their range of validity and critically evaluate the accuracy of the predictions.

Another very interesting mechanism is thermal rectification. This is the key for the realization of thermal diodes and may be relevant for applications. Thermal rectification has been
recently studied in electron \cite{craven, ruoko} and spin \cite{ala,vinitha1,vinitha2} systems.
When the on-site energy of the QDs are different the device has important rectification properties.
Exploring different parameters, we find that inverting the temperature gradient the magnitude of
the thermal current is reduced by more than an order of magnitude.

The paper is organized as follows. We present the basis of the theoretical description in Section \ref{model}. In different subsections we explain the model and its symmetry properties, a physical picture for the thermal transport and the equations for the current. The different methods used to calculate the heat current and thermal conductance are described in Section \ref{methods}. The results for a symmetric system ($\mu_L=\mu_R=$, $E_L=E_R$, and  $\Gamma_L=\Gamma_R$) 
are presented in Section \ref{res}. In Section \ref{recti} we 
calculate the rectification properties on the thermal current of an asymmetric system.
Section \ref{sum} is devoted to the summary and conclusions.

\section{Theoretical description}
\subsection{Model}
\label{model}

The Hamiltonian for the system sketched in Fig. \ref{esquema} reads
\begin{eqnarray}
H &=&\sum_{\nu }E_{\nu }d_{\nu }^{\dagger }d_{\nu }+Ud_{L}^{\dagger
}d_{L}d_{R}^{\dagger }d_{R}+\sum_{k\nu }\varepsilon _{k\nu }\,c_{k\nu
}^{\dagger }c_{k\nu }  \notag \\
&&+\sum_{k\nu \sigma }\left( V_{k\nu }\,c_{k\nu }^{\dagger }d_{\nu }+\text{%
H.c.}\right) ,  \label{ham}
\end{eqnarray}%
where  $\nu =L,R$ refers to the left and right dot or leads. The first term
describes the energy of an electron in each dot, the second term is the
Coulomb repulsion between electrons in different dots, the third term
corresponds to a continuum of extended states for each lead, 
and the last term is the hybridization
between electrons of each dot and the corresponding lead. 
In general, both leads are at different chemical potentials $\mu _{\nu }$
and temperatures $T_{\nu }$.  For most of the results presented here we
take $\mu _{\nu }=0$.

The couplings to the leads, assumed energy-independent, are expressed
in terms of the half width at half maximum of the spectral density in the
absence of the interaction

\begin{equation}
\Gamma _{\nu }=\pi \sum_{k}|V_{k\nu }|^{2}\delta (\omega -\varepsilon _{k\nu
}).  \label{del}
\end{equation}

Notice that this model is equivalent to a spinfull model, by identifying $L \rightarrow \uparrow$ and $R \rightarrow \downarrow$ (or vice versa) and considering
 fully polarized ferromagnetic leads with opposite orientation of the magnetization. \cite{pola} In such case, 
 electrons with a given spin orientation  can only tunnel to the lead polarized in the same direction and not to the other, as in the case of the Hamiltonian of Eq. (\ref{ham}).
 On the other hand, 
this model is also equivalent to the two-level model with destructive interference 
studied in Ref. \onlinecite{desint} under high magnetic field. In this case, the labels $L,R$ correspond to two different degenerate levels of the same dot and the continua that 
hybridizes with each of them.


\subsection{Transformations of the Hamiltonian}

\label{trans}

For later use, we describe here some transformations that map the Hamiltonian into itself with different parameters. Unless the parameters are left invariant by the transformations, they are not symmetries of the Hamiltonian. 
We assume that the details of the conduction bands are not important, and the corresponding spectral densities can be assumed constant.
Then, the electron-hole transformation 
\begin{eqnarray}
d_\nu^\dagger \rightarrow d_\nu,
c_{k\nu}^\dagger \rightarrow -c_{k^\prime \nu} \notag \\ 
\text{with } 
\varepsilon _{k^\prime \nu}=-\varepsilon _{k\nu }.
\label{eh}
\end{eqnarray}
except for an unimportant additive constant, leads to an Anderson model with the following transformed parameters
\begin{eqnarray}
E_\nu^\prime = - E_\nu - U,  \notag \\
\mu_\nu^\prime = - \mu_\nu, 
\label{ehp}
\end{eqnarray}
while $U$ and $\Gamma_{\nu}$ are unchanged.
Clearly this is a symmetry of the Hamiltonian in the so called {\em symmetric case} $\mu_\nu=0$, $E_\nu=-U/2$.

Similarly one could perform this transformation for only the left or right part of the Hamiltonian. 
In the latter case, the transformation is
\begin{eqnarray}
d_R^\dagger \rightarrow d_R,
c_{k R}^\dagger \rightarrow -c_{k^\prime R}.
\label{shiba}
\end{eqnarray}
In this case, the transformed parameters become
\begin{eqnarray}
U^\prime =-U, \notag \\
E_L^\prime =  E_L + U,  \notag \\
E_R^\prime = - E_R,  \notag \\
\mu_R^\prime = - \mu_R 
\label{shibap}
\end{eqnarray}
while the rest of the parameters remain unchanged.

In the symmetric case, this transformation maps the problem for Coulomb repulsion $U$ to a model with and attractive interaction $-U$.

\subsection{Mechanism for thermal transport}

\label{pict}

Since electrons cannot hop between left and right QDs, it is clear that the particle current is zero. 
It might seem surprising that the heat current is nonzero 
under a finite temperature difference $\Delta T= T_L-T_R$, in spite of the fact that the exchange of particles is not possible.
The aim of this section is to provide an intuitive picture for the transport of heat in the presence
of interactions. We assume small $\Gamma_\nu$ so that states with well-defined  number of particles
at each dot are relatively stable. Without loss of generality we can also assume $\Delta T>0$. 
Let us take $E_L=E_R < \mu_L=\mu_R=0$ and $E_\nu +U >0$. 

\begin{figure}[tbp]
\includegraphics[clip,width=7cm]{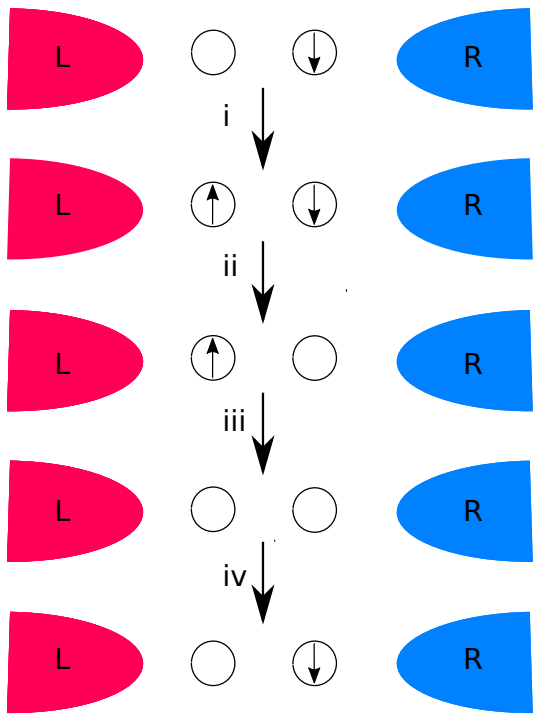}
\caption{(Color online) Schematic picture for the transport of heat in the presence of interactions. Although Eq. (\ref{ham}) is defined for spinless electrons, the sketch considers also
spin, in order to also represent  the corresponding configurations of the single-dot spinfull model with polarized reservoirs discussed at the end of Section \ref{model}. 
}
\label{esquema2}
\end{figure}

For $\Gamma_\nu \rightarrow 0$ one of the possible ground states of 
the system has occupancies $(n_L,n_R)=(0,1)$. Let us take this state as the initial state for a cycle of transitions 
that transport heat. For non-zero $\Gamma_L$, if $T_L$ is high enough there is a finite probability for an electron from the left lead 
to tunnel into the left QD and perform 
the thermal cycle  shown in Fig. \ref{esquema2}. The steps of the cycle are the following.
i) An electron from the left lead occupy the left dot changing the state of the double 
dot to (1,1) [(0,1) $\rightarrow$ (1,1)]. This costs energy $U+E_L$ which is taken from the left lead. 
Next (ii) an electron from the right dot hops to the right lead [(1,1) $\rightarrow$ (1,0)]. 
This relaxes the energy $U+E_R$ which is then transferred to the right lead. Next (iii) the electron from the left 
dot jumps to the corresponding lead [(1,0) $\rightarrow$ (0,0)]. This requires an energy $|E_L|$ taken form the left lead. 
Finally, (iv) an electron from the right lead occupy the right dot closing the cycle [(0,0) $\rightarrow$ (0,1)] 
and transferring the energy $|E_R|$ to the right lead. As a result of the cycle an amount of energy 
$U$ is transferred from the left to the right lead. 

The processes involved in each step of this cycle compete against the reverse  ones.
The resulting thermal current sketched in Fig. \ref{esquema2} actually
depends on the probability per unit time of these process
 and its proper evaluation requires an explicit calculation.
 In addition, while this picture provides a qualitative understanding 
for the general case, it is not enough to describe the thermal transport in the Kondo regime in which cotunneling events are important and does not explain what  happens in the $U \rightarrow \infty$ limit.

Fig. \ref{esquema2} is also useful to represent the fluctuations involved in the Kondo effect. For the spinfull QD coupled to polarized reservoirs mentioned at the end of Section \ref{model}
and for the model of Ref. \onlinecite{desint} these processes correspond to spin and orbital fluctuations, respectively. 
The sequence of the two steps (i) and (ii) and its time-reversed sequence
correspond to fluctuations through the virtual state with double occupancy. Note that a temperature difference $\Delta T>0$ favors the sequence (i)-(ii) with respect to the reciprocal one. Similarly, the process (iii)-(iv) and the reciprocal one correspond to fluctuations  through the virtual empty double dot and the former is favored by the temperature difference $\Delta T>0$.

We must warn the reader that the above simple picture uses eigenstates of the limit 
$\Gamma_\nu \rightarrow 0$ and does not explain the existence of a finite heat current 
for finite $E_\nu$ in the limit $U \rightarrow 0$ as discussed in Section \ref{uinf}. 

\subsection{Equations for the currents}

\label{curr}

We consider $T_R=T$ and $T_L=T+\Delta T$. The heat currents $J_{Q}^{L}$ flowing from the left lead to the dot and 
$J_{Q}^{R}$ flowing from the dot to the right lead are
\begin{equation}
J_{Q}^{\nu }=J_{E}^{\nu }-\mu _{\nu }J_{N}^{\nu },  \label{jq}
\end{equation}%
where $J_{E}^{\nu }$ are the energy currents.
In the stationary state, the charge and energy currents are uniform and should be conserved:
$J_N^L=J_N^R$ and $J_E^L=J_E^R$. The heat current is not conserved under an applied voltage 
($\mu_L \neq \mu_R$) due to Joule heating of the interacting part of the system.\cite{ng}
 For the setups studied in this work  $J_N^L=J_N^R=0$ because electrons cannot hop between the two QDs.  Hence, $J_N^{\nu}=0$ and the heat currents coincide with the energy currents. 

In terms of non-equilibrium Green's functions the latter are given by 
\begin{equation}
J_{E}^{\nu }=\pm \frac{2i\Gamma_{\nu }}{h}\int \omega d\omega \left[
2if_{v}(\omega )\text{Im}G_{\nu }^{r}(\omega )+G_{\nu }^{<}(\omega )\right] ,
\label{jelr}
\end{equation}
where upper (lower) sign corresponds to $\nu =L$ ($R$). The retarded  $G_{\nu }^{r}(\omega )$ and lesser $G_{\nu }^{<}(\omega )$ Green's functions are, respectively, the Fourier transforms of
$G_{\nu }^{r}(t-t^{\prime} ) = -i \Theta(t-t^{\prime}) \langle \{c_{\nu}(t), c^{\dagger}_{\nu}(t^{\prime}) \}\rangle$
and $G_{\nu }^{<}(t-t^{\prime} ) = i  \langle c^{\dagger}_{\nu}(t^{\prime})  c_{\nu}(t) \rangle$, while $f_{\nu }(\omega )=\left\{ 1+\exp
[(\omega -\mu _{\nu })/T_{\nu }]\right\} ^{-1}$ is the Fermi function.

In the next sections (particularly when the approach used conserves the heat current only approximately) we  adopt the following definition for  
the heat current  $J_Q=\sum_{\nu} J_E^{\nu}/2$.
For a small temperature difference, such that $\Delta T/T \ll 1$, the above equation  can be expanded  in powers of this quantity. The thermal conductance, is the coefficient associated to the linear order in this expansion. It is defined as
\begin{equation}\label{kappa}
\left. \kappa(T)=\frac{d J_Q}{d \Delta T}\right|_{\Delta T=0}.
\end{equation}

For completeness, we present below the expressions for the particle current flowing
between the  leads and the QDs in terms of Green's functions,
\begin{equation}
J_{N}^{\nu}=\pm \frac{2i\Gamma _{\nu}}{h}\int d\omega \left[ 2if_{\nu}(\omega )
\text{Im}G_{\nu}^{r}(\omega )+G_{\nu}^{<}(\omega )\right] . \label{jnu}
\end{equation}
In the next section, we will use the fact that $J_N^{\nu}=0$, in order to infer properties of the Green's functions in the limit $\Gamma_{\nu}\rightarrow 0$.

\section{Methods}
\label{methods}

We now briefly describe the methods to be used to calculate the heat current in the different regimes of parameters.

\subsection{Weak coupling to the reservoirs}
\label{ato}

This limit corresponds to $\Gamma_{\nu} \rightarrow 0$. This regime is usually addressed with rate equations.\cite{rafa1,rafa2,ruoko,yada}
Here, we present an alternative derivation on the basis of Green's functions and conservation laws. 
Evaluation of Eq. (\ref{jelr}) at the lowest order in $\Gamma_{\nu}$ implies calculating $G_{\nu}^{r,<}(\omega )$ for the QDs uncoupled from the reservoirs. This is usually referred to as  the {\em atomic limit}, and
the Green's functions can be calculated, for instance, from equations of motion. \cite{none} The result is
\begin{eqnarray}
G_{\nu }^{r}(\omega ) &=&\frac{1-n_{\bar{\nu}}}{\omega -E_{\nu }}+\frac{n_{%
\bar{\nu}}}{\omega -E_{\nu }-U},  \notag \\
G_{\nu }^{<}(\omega ) &=&2\pi i\left[ a_{\nu }\delta (\omega -E_{\nu
})+b_{\nu }\delta (\omega -E_{\nu }-U)\right] ,  \label{gf}
\end{eqnarray}%
where $n_{\nu }=\langle d_{\nu }^{\dagger }d_{\nu }\rangle $ is the
expectation value of the occupancy of the dot $\nu $ and $\bar{\nu}=R$ ($L$)
if $\nu =L$ ($R$). The functions $a_{\nu }$ and $b_{\nu }$ are not simply determined by the energy population of the reservoirs
for $\Gamma _{\nu }=0$. Within the equation of motion technique it is necessary 
 to include finite $\Gamma _{\nu }$ and approximations in order to evaluate them. Another possibility is to evaluate them from rate equations. \cite{rafa1,rafa2,ruoko} Here we proceed as follows. We start by 
 expressing the occupation of the QD in terms of Green's functions
\begin{equation}
n_{\nu }=\frac{-i}{2\pi }\int d\omega G_{\nu }^{<}(\omega )=a_{\nu }+b_{\nu
}.  \label{nnu}
\end{equation}
The latter equation defines a relation between the occupation and the unknowns $a_{\nu }$ and $b_{\nu}$.

Replacing Eqs. (\ref{gf}) and (\ref{nnu}) in Eqs. (\ref{jnu})
and imposing $J_{N}^{L}=J_{N}^{R}=0$ we get the following set of two
equations 

\begin{equation}
n_{\nu }=(1-n_{\bar{\nu}})f_{\nu }(E_{\nu })+n_{\bar{\nu}}f_{\nu }(E_{\nu
}+U),  \label{seq}
\end{equation}%
from which $n_{\nu }$ can be determined. The result is 

\begin{eqnarray}
n_{\nu } &=&\frac{f_{\nu }(E_{\nu })-f_{\bar{\nu}}(E_{\bar{\nu}})D_{\nu }}{%
1-D_{L}D_{R}},  \notag \\
D_{\nu } &=&f_{\nu }(E_{\nu })-f_{\nu }(E_{\nu }+U).  \label{nnuf}
\end{eqnarray}

Using Eqs. (\ref{seq}) the energy currents can be written as

\begin{equation}
J_{E}^{\nu }=\pm \frac{4\pi \Gamma _{\nu }U}{h}\left[ n_{\bar{\nu}}f_{\nu
}(E_{\nu }+U)-b_{\nu }\right] .  \label{je2}
\end{equation}%
Conservation of the energy current in the stationary state $%
J_{E}^{L}=J_{E}^{R}$ leads to an equation for $\Gamma _{L}b_{L}+\Gamma_{R}b_{R}$. At this point we introduce the assumption $b_{L}=b_{R}$. This is
justified from the functional dependence of  $G_{\nu }^{<}(\omega )$ on these parameters [see Eqs. (\ref{gf})] and Eq.(\ref{nnu}). In fact, notice that 
$b_{\nu}$ is the contribution to $n_{\nu}$ at the
energy $E_{\nu}+U$, which implies that the two dots are occupied. Hence, 
$b_{\nu }=\langle d_{L}^{\dagger }d_{L}d_{R}^{\dagger }d_{R}\rangle $ for $\nu=L,R$.

Therefore, using $J_{E}^{L}=J_{E}^{R}$ and $b_{L}=b_{R}$ we obtain

\begin{eqnarray}
(\Gamma _{L}+\Gamma _{R})b_{\nu } &=&\Gamma_{L}n_{R}f_{L}(E_{L}+U)  \notag\\
&&
+\Gamma _{R}n_{L}f_{R}(E_{R}+U).  \label{bnu}
\end{eqnarray}%
Using Eqs. (\ref{nnuf}) and some algebra we can verify that Eq.  (\ref{bnu}) leads to the correct result at equilibrium. In fact,
 for $\mu _{L}=\mu_{R}=0$, $T_{L}=T_{R}=1/\beta $ we recover

\begin{eqnarray}
b_{\nu } &=&\langle d_{L}^{\dagger }d_{L}d_{R}^{\dagger }d_{R}\rangle
=n_{R}f_{L}(E_{L}+U)=n_{L}f_{R}(E_{R}+U)  \notag \\
&=&\frac{e^{-\beta (E_{L}+E_{R}+U)}}{1+e^{-\beta E_{L}}+e^{-\beta
E_{R}}+e^{-\beta (E_{L}+E_{R}+U)}}.  \label{beq}
\end{eqnarray}

Replacing Eq. (\ref{bnu}) in Eq. (\ref{je2}) we obtain the final
expression for the heat current,

\begin{eqnarray}
J_{Q} &=&\Gamma_Q U [n_{R}f_{L}(E_{L}+U)  
-n_{L}f_{R}(E_{R}+U)], \label{jqf}
\end{eqnarray}
with $\Gamma_Q= 4\pi \Gamma _{L}\Gamma _{R}/h(\Gamma_{L}+
\Gamma_{R})$.

After some algebra, it can be checked that this expression is invariant under the transformations defined by
Eqs. (\ref{ehp}) and (\ref{shibap}), as expected.

For the symmetric case $\mu_\nu=0$, $E_{\nu }=-U/2$, we have $n_{\nu }=1/2$ and Eq. (\ref{jqf})
reduces to 
\begin{eqnarray}
J_{Q} &=& \frac{\Gamma_Q}{2}U [f_{L}(U/2)-f_{R}(U/2)]. \label{jqyada}
\end{eqnarray} 
which coincides with the expression obtained by Yadalam and Harbola [see the
expression of $C_{1}$ in appendix A of Ref. \onlinecite{yada}, note that 
in their notation $\Gamma_\nu$ is two times our definition given by Eq. (\ref{del})].
Interestingly, in this case, the symmetry Eq. (\ref{shiba}) implies that $J_Q$ is an even function 
of $U$.

\subsection{Moderate or strong coupling to the reservoirs}
\label{scoup}

We briefly introduce the  methods we use to solve the problem for finite $\Gamma_\nu$, 
discussing the range of validity, as well as
the advantages and disadvantages. These are perturbation theory (PT), renormalized perturbation theory (RPT) and non-crossing approximation (NCA). All these methods are suitable to address the Kondo regime.

\subsubsection{Perturbation theory in $U/\Gamma$ (PT)}
\label{pert}

For the Anderson model at equilibrium, with $\Gamma_L=\Gamma_R=\Gamma$,  PT 
in the Coulomb repulsion $U$
has been a popular method used for several years now,\cite{yam,hor}
also applied to nanoscopic systems at equilibrium as well as away from equilibrium,\cite{levy,ogu,mir,pro,lady,hersh,none} and recently to superconducting 
systems.\cite{zonda,zonda2} It consists in calculating the Green's function with a self-energy evaluated up to second order in the interaction $U$.
As expected for a perturbative approach, it is in principle valid for $U/(\pi \Gamma) <1$. However 
comparison with Quantum Monte Carlo results indicate that the method in equilibrium configurations
is quantitatively valid in the symmetric case $E_L=E_R=-U/2$, for $U/(\pi \Gamma )$ as large as 
2.42.\cite{sil} 

Here we summarize the main expressions  following the notation of Ref. \onlinecite{none}. The retarded and lesser Green's functions read
\begin{eqnarray}\label{2o}
\left[G^r_{\nu}(\omega)\right]^{-1}&=& \left[g^r_{\nu}(\omega)\right]^{\-1} -\Sigma_\nu^{r2}(\omega),\nonumber \\
G^<_{\nu}(\omega) &=& |G^r_{\nu}(\omega)|^2 \left(\frac{g^<_{\nu}(\omega)}{|g^r_{\nu}(\omega)|^2} - \Sigma_\nu^{<2}(\omega)\right).
\end{eqnarray}
The latter depend on the non-interacting Green's functions for the QDs coupled to the reservoirs,
\begin{eqnarray}
\left[g^r_{\nu}(\omega)\right]^{-1} &=& \omega - \epsilon_{\nu} +i \Gamma_{\nu}, \nonumber \\
g^<_{\nu}(\omega) &=&2 i |g^r_{\nu}(\omega)|^2 \Gamma_{\nu} f_{\nu}(\omega),
\end{eqnarray}
where $\epsilon_{\nu}$ are effective energies that contain the first-order corrections in $U$.
They vanish in the symmetric case $\mu_{\nu}=0$, $E_\nu=-U/2$.
The second-order contributions to the self-energies are

\begin{eqnarray}
\Sigma _{\nu}^{r2}(\omega ) &=&U^{2}\int \frac{d\omega _{1}}{2\pi }
\int \frac{d\omega _{2}}{2\pi }  \nonumber \\
&&[g_{\nu}^{r}(\omega _{1})g_{\bar{\nu}}^{r}(\omega
_{2})g_{\bar{\nu}}^{<}(\omega _{1}+\omega _{2}-\omega )  \nonumber \\
&&+g_{\nu}^{r}(\omega _{1})g_{\bar{\nu}}^{<}(\omega
_{2})g_{\bar{\nu}}^{<}(\omega _{1}+\omega _{2}-\omega )  \nonumber \\
&&+g_{\nu}^{<}(\omega _{1})g_{\bar{\nu}}^{r}(\omega
_{2})g_{\bar{\nu}}^{<}(\omega _{1}+\omega _{2}-\omega )  \nonumber \\
&&+g_{\nu}^{<}(\omega _{1})g_{\bar{\nu}}^{<}(\omega
_{2})g_{\bar{\nu}}^{a}(\omega _{1}+\omega _{2}-\omega )],  \label{sr2}
\end{eqnarray}

\begin{eqnarray}
\Sigma _{\nu}^{<2}(\omega ) &=&-U^{2}\int \frac{d\omega _{1}}{2\pi }
\int \frac{d\omega _{2}}{2\pi }  \nonumber \\
&&g_{\nu}^{<}(\omega _{1})g_{\bar{\nu}}^{<}(\omega
_{2})g_{\bar{\nu}}^{>}(\omega _{1}+\omega _{2}-\omega ),  \label{sl2}
\end{eqnarray}
where $g_{\nu}^{a}=\bar{g_{\nu}^{r}}$ is the advanced non-interacting Green's function.

Some integrals can be calculated analytically as described in Ref. \onlinecite{none}.

One shortcoming of the approach is that it does not guarantee the conservation of the particle and energy currents.
This means that in general the approximation gives $J_N^L \ne J_N^R$  and $J_E^L \ne J_E^R$ 
[see Eqs. (\ref{jelr}) and (\ref{jnu})], contrary to what one expects.
In our case however $J_N^L=J_N^R=0$ within numerical precision, so that the particle current is conserved. Concerning the energy current,
the relative deviation 
\begin{equation}\label{d}
d=|J_Q^L/J_Q-1|=|J_Q^R/J_Q-1|,
\end{equation}
 is usually of the order of 2\% or less,
but reaches a value near 14\% at high temperatures and the largest values of $U$ considered
with this method in Section IV ($U=7 \Gamma$). In the calculations, we will define the range of validity of the method as that corresponding to a small value of this deviation.

\subsubsection{Renormalized perturbation theory (RPT)}

\label{rpt}

For $U \gg \Gamma$ the approach mentioned above fails. However, for energy scales below 
$T_K$ one can use renormalized perturbation theory (RPT). 
For energy scales of the order of $T_K$ or larger, 
the method loses accuracy and a complementary approach is needed.

The basic idea of RPT is to calculate the Green's functions of  
Eq. (\ref{2o}) with the second-order self-energy calculated with renormalized
parameters 
$\widetilde{U}/(\pi \widetilde{\Gamma })$. The latter correspond to 
fully dressed quasiparticles, taking as a basis the equilibrium Fermi liquid picture.\cite{he1} 
The renormalized parameters  can be
calculated exactly using Bethe ansatz, or with high accuracy 
using numerical renormalization group.\cite{cb,re1} The resulting values of $\widetilde{U}/(\pi \widetilde{\Gamma })$ are small, being usually below 1.1
even for $U \rightarrow \infty$.\cite{cb,ct}

Our RPT procedure consists in using renormalized parameters for $E_L=E_R$, $U$ and $\Gamma $
obtained at $\mu_L=\mu_R=T_L=T_R=0$ by a numerical-renormalization-group calculation,\cite{cb,ct} 
and incorporating perturbations up to second order in the
renormalized $U$ ($\widetilde{U}$). It has been shown explicitly that this  satisfies 
important Ward identities even away from equilibrium.\cite{ng,ct}
At equilibrium, the method provides results that coincide
with state-of-the art techniques for the dependence of the electrical conductance with
magnetic field \cite{cb} and temperature \cite{ct}.

As in the case of PT, the conservation laws are not guaranteed and the relative deviation defined in Eq. (\ref{d}) is a control parameter to define the validity of the results 
obtained with this method.

\subsubsection{Non-crossing approximation (NCA)}

\label{nca}

The NCA technique is one of the standard tools for calculating these Green functions
in the Kondo regime, where the total occupancy of the interacting subsystem 
is near 1 and with small charge fluctuations 
(the charge is well localized in the dot or dots). It corresponds to evaluating the self-energy for the non-equilibrium Green's functions entering Eqs. (\ref{jelr}) and (\ref{jnu})
via the summation of an infinite series of diagrams (all those in which the propagators do not cross) in perturbation theory in the couplings $\Gamma_\nu$.\cite{win,roura_1,tosib}
The formalism is explained in detail in Ref. \onlinecite{tosib} for a model which 
contains the present one as a limiting case. 
The main limitation of the approach is that if fails to reproduce Fermi-liquid properties at 
equilibrium at temperatures well below the characteristic Kondo temperature.\cite{win}

NCA has being
successfully applied to the study of a variety of systems such as 
C$_{60}$ molecules displaying a quantum phase transition,\cite{serge,roura_2}, 
a nanoscale Si transistor,\cite{tetta}
two-level quantum dots,\cite{tosi_1} and the interplay between vibronic effects and the Kondo effect.\cite{desint,sate}  
Recently it has also been used to calculate heat transport.\cite{dare1,diego,dare2}
In spite of this success, the NCA has some
limitations at very low temperatures (below $\sim 0.1 T_{K}$). For example, 
it does not satisfy accurately the Friedel sum rule at zero
temperature.\cite{fcm}
In this sense it is complementary to RPT, which should be accurate for $T_L, T_R \ll T_K$.

In contrast to the previous methods, NCA conserves the charge current, as shown explicitly  in Ref. \onlinecite{tosib}. 
We find that the NCA also conserves the energy current.

\subsection{Evaluation of the Kondo temperature}

\label{tk}

In order to compare the results of different approximations, it is convenient to represent the results 
taking the unit of energy as the Kondo temperature $T_K$ which is the only relevant energy scale at small temperatures. 
The evaluation of $T_K$ in the Anderson impurity model on the basis of PT, RPT and NCA has been addressed in several works in the literature. Because of some details of the different approximations (like the high-energy 
cutoff for example), the $T_K$ differ, although they are of the same 
order of magnitude.

Here,
we follow Ref. \onlinecite{asym}, which is based on the analysis of the electrical conductance 
as a function of temperature $G(T)$ of a similar physical system. In the model under investigation, the electrical conductance vanishes, as already mentioned. However, it is possible to  define an Anderson impurity model, equivalent to ours at equilibrium,  
with non-vanishing $G(T)$,
which has the same $T_K$ as the model of Eq. (\ref{ham}). Recalling that the hybridization to the leads  is given by $\Gamma_L=\Gamma_R=\Gamma$,
the equivalent spin-degenerate Anderson  impurity  model has
hybridization $\Gamma/2$ for each spin orientation and each lead (this is actually the simplest Anderson model that describes the conductance through a spin degenerate QD).  
At equilibrium, the latter has the same spectral density and the same $T_K$ as our 2-dot model of Eq. (\ref{ham}). Hence, we can use the method of Ref. \onlinecite{asym} to calculate $T_K$.
In particular, we can extract $T_K$ from the temperature dependence of the electrical conductance $G(T)$. As explained in that reference, such a procedure is more reliable than alternative methods, like fitting  the spectral density or the non-equilibrium electrical conductance $G(V_b)$ 
as a function of bias voltage $V_b=(\mu_L-\mu_R)/e$. 
For example, different fitting procedures to fit the line shape of $G(V_b)$ 
of the same physical system differ by a factor two.\cite{asym}

Concretely, we fitted  a popular phenomenological expression for $G(T)$ 
for the parameters used in Section \ref{res} with $U \rightarrow \infty$.  
The renormalized parameters for RPT were taken from previous calculations.\cite{cb,ct} The result is
$T_K=0.00441 \Gamma$ for the RPT and $T_K=0.00796 \Gamma$ for the NCA.
In the symmetric cases with $U= 7 \Gamma$ and $U= 4 \Gamma$, we have obtained $T_K$ from the condition
$G(T_K)=G(T=0)/2$ with the result $T_K=0.191 \Gamma$ for $U= 7 \Gamma$ and $T_K=0.495 \Gamma$ 
for $U= 4 \Gamma$.

\section{Results for a symmetric device}
\label{res}

In this section we analyze and compare results for the thermal response of the system under investigation, calculated with the different methods presented in the previous section.
All calculations in this Section were done for $\mu_L=\mu_R=0$, $E_L=E_R=E$, and 
$\Gamma_L=\Gamma_R=\Gamma$. In this case the system has reflection symmetry under a plane 
bisecting the device.

\subsection{Thermal conductance}
\label{tcond}

We start by analyzing the behavior of the thermal conductance defined by Eq. (\ref{kappa}). For weak coupling to the reservoirs, and in the symmetric case $E_\nu=-U/2$,  we can expand 
 Eq. (\ref{jqyada}) up to linear order in  $\Delta T$ and we get the following result,
\begin{equation}\label{kappa-weak}
\kappa(T)= \frac{\Gamma_Q}{4 k_B}  \; e^{U/2k_B T}\left[\frac{U}{T} f_R(U/2)\right]^2.
\end{equation}
This expression is exact in the limit $\Gamma \ll k_B T, U$. 

\begin{figure}[h!]
\begin{center}
\includegraphics[clip,width=\columnwidth]{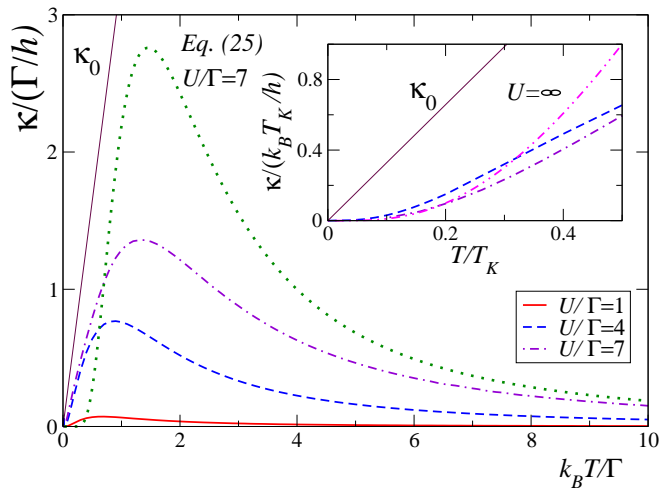}
\caption{(Color online) Main panel: Thermal conductance as a function of the temperature for the symmetric configuration $E_L=E_R=-U/2$ and different values of $U$, calculated with PT (see 
Section \ref{pert}). The dotted line 
indicates the weak-coupling prediction Eq. (\ref{kappa-weak}) for $U/\Gamma=7$. The thin line corresponds to the quantum of thermal conductance [Eq. (\ref{quantum})].
 Inset:  Low-temperature part of the plots for $U/\Gamma=4, 7$, with the temperature expressed in units of $T_K$. The limit $U \rightarrow \infty$, calculated with RPT is also shown by the 
dash-dot-dot line.}
\label{figkthper}
\end{center}
\end{figure}

Results for strong-coupling to the reservoirs  are shown in Fig. \ref{figkthper}. We consider $\Gamma_L=\Gamma_R=\Gamma$, $\mu_L=\mu_R=0$. The results of the figure are calculated with the perturbation theory presented in Section \ref{pert}. The values of the Coulomb interaction $U$ have been chosen within the range
where the method has been probed to be reliable for the description of the electrical conductance. \cite{ogu} We have also verified that, for these values, the conservation of the energy current is satisfied within 
and error of $14\%$ at the most. For $U/\Gamma=7$, we also include the prediction of Eq. (\ref{kappa-weak}) (see plot in dotted line in the main panel). The latter tends to coincide with the results of PT in the high-temperature limit. 
In Fig. \ref{figkthper} we also indicate the reference defined by the quantum of thermal 
conductance $\kappa_0$ [see Eq. (\ref{quantum})], which defines an upper bound per transmission channel.  At low temperatures, $T\ll T_K$,
we find by a numerical fit of the results that $\kappa \sim T^4$. For $T \sim 0.5 T_K$ the dependence on $T$ evolves to linear. The  slope  of $\kappa$ for $T < \Gamma$ becomes closer to the slope of 
$\kappa_0$ as $U$ increases. At temperatures $k_B T \sim \Gamma$ 
the conductance achieves a maximum and then decreases exponentially for larger $T$.  The low-temperature regime can be further analyzed by representing the thermal conductance as a function of $T/ T_K$ (see inset of the Fig \ref{figkthper}). 
The method used to determine $T_K$ is discussed in Section \ref{tk}.
The limiting case of $U\rightarrow \infty$, calculated with RPT (valid for $T \ll T_K$), is also shown. These results suggest a universal behavior of $\kappa$, independent of $U$, deep in the Kondo regime at small temperatures. Some deviations are however noticeable for $U=4 \Gamma$ for which charge fluctuations are very important, and the system is not strictly in the Kondo regime 
$-E_\nu, E_\nu+U \gg \Gamma$.

\begin{figure}[h!]
\begin{center}
\includegraphics[clip,width=\columnwidth]{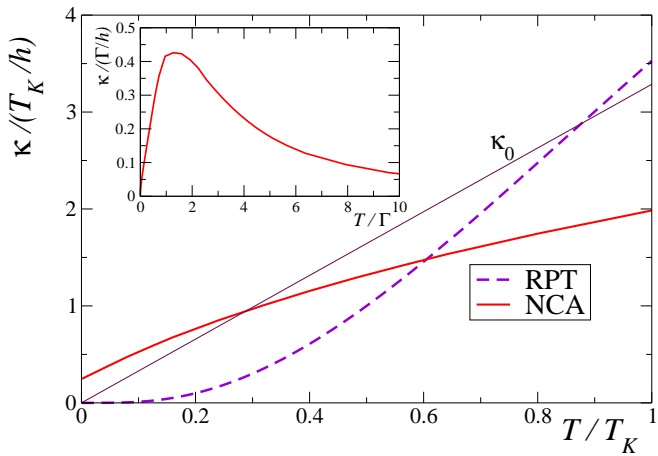}
\caption{(Color online) Main panel: Low-temperature behavior of the thermal conductance in the limit $U \rightarrow \infty$ for $E_L=E_R=-4\Gamma$, as a function of $T/T_K$, calculated with RPT (see Section \ref{rpt}) and NCA (see section \ref{nca}). The straight line corresponds to the quantum
bound [Eq. (\ref{quantum})]. Inset:  Thermal conductance as a function of $T$ calculated with NCA. 
}
\label{figkthuinf}
\end{center}
\end{figure}

In Fig. \ref{figkthuinf} we present results for the thermal conductance in the limit $U \rightarrow \infty$ for temperatures  below $T_K$.  We compare the results obtained with NCA 
(see Section \ref{nca}) and RPT (Section \ref{rpt}). In this case, the parameters do not correspond
to the symmetric configuration. However, they correspond to the Kondo regime $-E_\nu, E_\nu+U \gg \Gamma$.

NCA overestimates the thermal conductance at the lowest temperatures, leading to a prediction higher than the upper bound 
$\kappa_0$. This method is known to fail the description of the electrical conductance at very low $T$.\cite{win,asym} Here, we see that it is also inadequate to predict the low-temperature behavior of the thermal conductance.
 In the case of RPT, the conservation of the energy current is satisfied to a good degree for 
 $T \leq 0.4 T_K$ with  $d <5\%$ [see Eq. (\ref{d})]. This suggests the validity of this method to calculate the thermal conductance at low temperature. 
As in the symmetric case shown in the inset of Fig. \ref{figkthper}, the  behavior  is consistent with a power law $\kappa \sim T^4$.  
RPT overestimates the response at higher temperatures, close to $T_K$, where it is not expected to be valid.\cite{ng} 
The fact that the RPT result for $\kappa (T)$ is above the lower bound given by Eq. (\ref{quantum})
for $T=T_K$ clearly shows the breakdown of the approximation at this temperature. In fact,
for increasing temperature, the error in the conservation of the current increases.
Specifically, the relative deviation $d$ is below 1\% for $T < 0.23 T_K$, increases to 7.7\% for
$T=T_K/2$ and to nearly 17\% for $T=T_K$. Overall, the analysis of these results suggests that deep in the Kondo regime, $T\ll T_K$, RPT is very likely to predict the correct behavior of $\kappa(T)$, while
for $T > T_K/2$, the NCA results are more reliable. Therefore, as in the case of the electric current,\cite{asym} both approaches are complementary. 
It is encouraging to see that 
in the transition between the range of validity of both approaches, they give the same order of magnitude of the thermal conductance when the results are scaled by the corresponding $T_K$.

The important physical outcome of this analysis is that we find a significant enhancement of the thermal response deep in the Kondo regime, in relation to the limit of weak coupling to the reservoirs.
Concretely, $\kappa \sim T^{4}$, for $T \ll T_K$ and $\kappa \propto T$ for $T\sim T_K$, with a proportionality constant smaller to the one in the quantum bound of Eq. (\ref{quantum}). This is in contrast  
with the exponentially small thermal conductance at low temperatures given by Eq. (\ref{kappa-weak}), in the case of very low coupling to the reservoirs.

\subsection{Far-from-equilibrium thermal response}
The aim of the present section is to analyze the thermal current when $\Delta T= T_L-T_R >0$. As in the previous section, we consider $\Gamma_L=\Gamma_R=\Gamma$, $\mu_L=\mu_R=0$. 

\subsubsection{Dependence of thermal current on $\Gamma_\nu$}
\label{depdelta}

\begin{figure}[h!]
\begin{center}
\includegraphics[clip,width=\columnwidth]{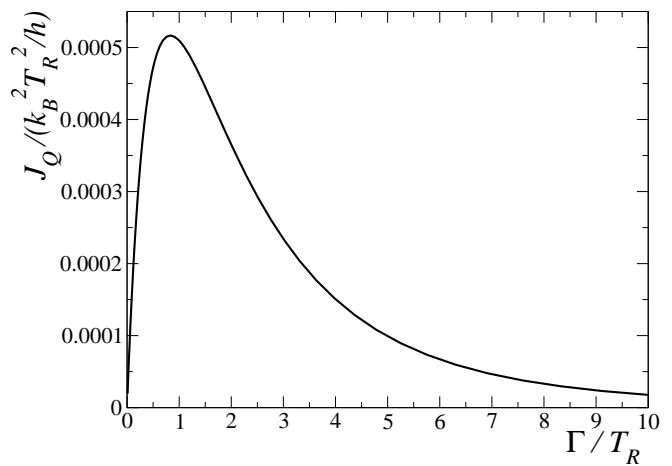}
\caption{Thermal current as a function of dot-lead couplings For $T_L=2T_R$, 
$U=k_BT_R/10$ and $E_L=E_R=-U/2$. }
\label{figdelta}
\end{center}
\end{figure}

We start by presenting in Fig. \ref{figdelta} results for small  $U=k_BT_R/10$, where $T_R$ is the temperature of the coldest reservoir, in the symmetric configuration where $E_L=E_R=-U/2$. These parameters correspond to the high-temperature regime where the Kondo effect is not developed in equilibrium.
The evaluation have been done with perturbation theory as explained in Section \ref{pert}, which is accurate within the range $\Gamma \gg U$.  
In the weak-coupling limit, with $\Gamma < U$ the thermal current is given by Eq. (\ref{jqf}), which for the symmetric configuration reduces to Eq. (\ref{jqyada}).
In the description of PT, the heat current increases linearly with $\Gamma$ for small values of these parameters as in Eq. (\ref{jqyada}). 
For larger 
$\Gamma$, the slope decreases and $J_Q$ reaches a maximum for $\Gamma \sim 0.8 T_R$. Then, it decreases for larger $\Gamma$. For all values of $\Gamma$ the relative deviation of the method in the conservation of the energy current $d$  is below 1\%. Interestingly, the qualitative behavior of $J_Q$ calculated with PT is very similar to the one presented  in Ref. \onlinecite{yada}, on the basis of a saddle point approximation within the path-integral formalism. 


\subsubsection{Dependence of the thermal current on $\Delta T$}

\label{dept}

\begin{figure}[h!]
\begin{center}
\includegraphics[clip,width=7.5cm]{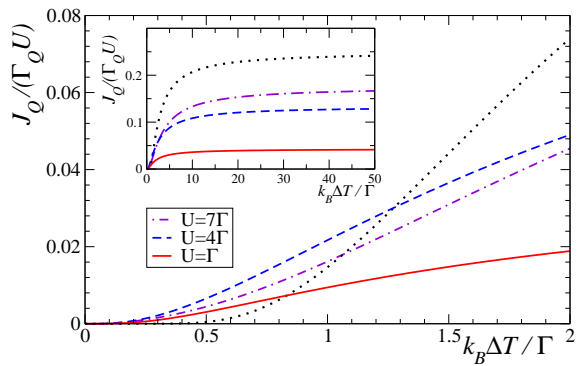}\\
\caption{(Color online) Thermal current as a function of $T_L=\Delta T$ for $T_R=0$, 
$E_L=E_R=-U/2$ and several values of $U$. The dotted line corresponds to Eq. (\ref{jqyada}) with $U=7\Gamma$.}
\label{figt}
\end{center}
\end{figure}

We now take $T_R=0$ and analyze the dependence of $J_Q$ on $T_L=\Delta T$,
using perturbation theory in $U$ for the symmetric case $E=E_L=E_R=-U/2$, as above. 
We consider several values of $U$ within the validity of  PT.
The results are shown in Fig. \ref{figt}. As in the previous section, we evaluate the limits of this
approach from the relative error in the conservation of the energy current 
$d$. We have verified that it is negligible for very small temperatures 
and moderate values of $U$, while it reaches a value of 12.6 \% for $U=7 \Gamma$ and $T_L \sim 2 \Gamma$, decreasing slowly with further increase in $T_L$.

For $U=7 \Gamma$, the system is in the Kondo regime ($-E, E+U \gg \Gamma$) at low temperatures and at equilibrium. Correspondingly, the spectral 
density has a well defined peak at the Fermi energy (the Kondo peak) separated from the charge-transfer peaks
near $E$ and $E+U$. 
For $T_L$ well below $T_K$ (we verify this for $T_L < 0.04 \Gamma$), the heat current behaves as $J_Q \sim (\Delta T)^4$. 
This remains true as long as the smaller temperature ($T_R$ in our case) is also much smaller than  
$T_K$. For large $T_R$, $J_Q$ is linear in $\Delta T$ for small $\Delta T$. 

For all values of $U$, after the initial slow increase of the thermal current with $\Delta T$, for 
$\Delta T \sim \Gamma$, $J_Q$ increases approximately linearly with $\Delta T$ and when it reaches a few times $U$ if finally saturates. For $U, T_R, \Delta T \gg \Gamma$, the thermal current is described by Eq. (\ref{jqyada}).
If in addition $T_L, T_R \gg U$, the heat current in the weak-coupling limit behaves as
\begin{equation}\label{bound}
J_Q \simeq \frac{\Gamma_Q U}{8}  \; \left[ U \left(\frac{1}{T_R}-\frac{1}{T_L} \right) +\frac{E_L}{T_L}-\frac{E_R}{T_R} \right],
\end{equation}
which in the limit of $\Delta T \gg T_R$ saturates to $J_Q \sim \Gamma_Q U(U-E_R)/T_R $.  

The result in the atomic limit described by Eq. (\ref{jqyada}), is shown in dotted line in the inset 
for $U= 7 \Gamma$. 
Note that the saturation value of Eq. (\ref{jqyada}) for $T_R=0$ at high $T_L$ is  $\Gamma_Q U/4$.
Therefore, for the units of $J_Q$ chosen, the curves for different $U$ coincide at large $\Delta T$.
Clearly only for $U \gg \Gamma$ the saturation value of $J_Q$ for large $\Delta T$ predicted by RPT approaches the corresponding value in the atomic limit.

As in the case of the thermal conductance,  when the two reservoirs have temperatures $T_R, T_L < T_K$, there is a strong enhancement in the value of the thermal current, for dots strongly coupled to reservoirs relative to the case where they are weakly coupled.  In fact, the current is exponentially small  at low temperatures for weakly coupled quantum dots. Instead, 
within the Kondo regime, $J_Q \propto (\Delta T)^4$ for $T_R, T_L  \ll T_K$, while  $J_Q \propto (\Delta T) $ if either $\Delta T > T_K/2$ or the temperature of the coldest lead $T_R > \Delta T$. 
Note that the latter case corresponds to the calculation of the thermal conductance defined by 
Eq. (\ref{kappa}).

\subsubsection{The limit $U \rightarrow \infty$}

\label{uinf}

Here we choose parameters corresponding to the Kondo regime 
(defined by $-E_\nu, E_\nu+U \gg \Gamma$) at equilibrium: $E_L=E_R=E=-4 \Gamma$, and 
$U \rightarrow \infty$, and calculate the current using RPT and NCA (see Sections \ref{rpt}, \ref{nca}) as a function of $\Delta T$, keeping $T_R=0$ (RPT) or a small fraction of the 
Kondo temperature (NCA) so that the results are indistinguishable from those of $T_R=0$ at equilibrium.

As in Section \ref{tcond}, in order to compare the results of RPT and NCA we scale the properties by
the corresponding Kondo temperature obtained previously,\cite{asym} as explained in that Section.

\begin{figure}[h!]
\begin{center}
\includegraphics[clip,width=\columnwidth]{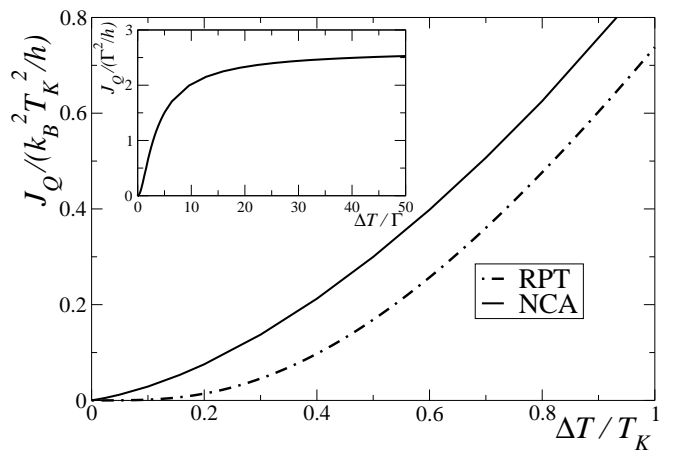}
\caption{Thermal current as a function of $\Delta T$ for  $U \rightarrow \infty$  
and $E=-4 \Gamma$. Inset: Thermal current calculated with NCA for a wide range of temperatures.}
\label{figuinf}
\end{center}
\end{figure}

The result for $J_Q$ as a function of $\Delta T$ is shown in Fig. \ref{figuinf}. For small
temperatures $T < 0.2 T_K$, the RPT result is more reliable and shows a dependence $J_Q \sim \Delta T^4$. 
For $T > T_K$, the RPT breaks down. 
We show in the inset of Fig. \ref{figuinf} only results calculated with NCA for a wide range of temperatures, including the
high-temperature regime $T \gg T_K$. 
In the latter plot, we observe that for  high temperatures, 
the thermal current saturates to a finite value. This behavior is the same observed for finite $U$  in the symmetric case (see Fig. \ref{figt}).
Notice that in the limit of $\Gamma \rightarrow 0$, Eq. (\ref{jqf}) $J_Q \rightarrow 0$ for $U \rightarrow \infty$. Instead, the present results show that the thermal current 
at finite coupling to the reservoirs is ${\cal O}(\Gamma^2)$.   Eq. (\ref{jqf}) does not account for this contribution, since it
 corresponds to the \textit{linear} order term in $\Gamma$ of the thermal current.

\begin{figure}[h!]
\begin{center}
\includegraphics[clip,width=7cm]{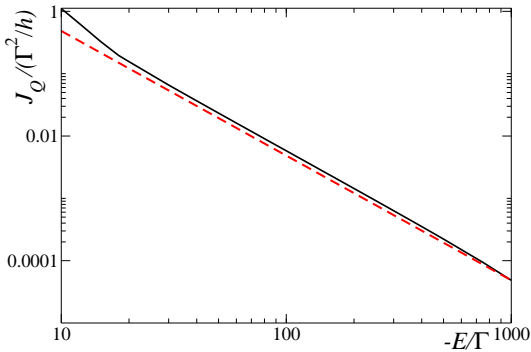}
\caption{(Color online) Full line: Thermal current as a function of dot energies $E$ for $k_B \Delta T =50 \Gamma$ and  $U \rightarrow \infty$. Dashed line corresponds to $A/E^2$ where $A$ is a constant.}
\label{jvse}
\end{center}
\end{figure}

The above analysis has been done at a finite $E$, and one might wonder if a finite current remains 
for $U \rightarrow \infty$ keeping $E=-U/2$ (symmetric case). Because of technical reasons, we can not directly address this question with the methods used in the present section.
We can in any case gather some intuition by calculating the dependence of the 
thermal current with $E$ within NCA at a  high temperature $T \gg T_K$.
The result is shown in Fig. \ref{jvse}. The current decreases with increasing $-E$.
We find that the dependence with the energy levels of both dots (taken equal) is very near  
$E^{-2}$ for $E \gg \Delta T$. This result indicates that the thermal current vanishes for 
$U \rightarrow \infty$ in the symmetric case $E=-U/2$. The reason for such a different behavior between the symmetric and non-symmetric configuration is that in the former one, there is a vanishing spectral weight at the Fermi energy. 
In fact, for the far-from equilibrium situation analyzed here, the Kondo peak that develops at equilibrium is completed melted and the spectrum consists of the two  Coulomb blockade peaks, which for $U \rightarrow \infty$ are at an infinitely high energy.
Instead, 
for the parameters of Fig. \ref{figuinf}, there is some finite spectral weight at energies $E_L=E_R$, which enable the thermal transport.

\subsubsection{Dependence of thermal current on $U$}

\label{depu}

\begin{figure}[h!]
\begin{center}
\includegraphics[clip,width=\columnwidth]{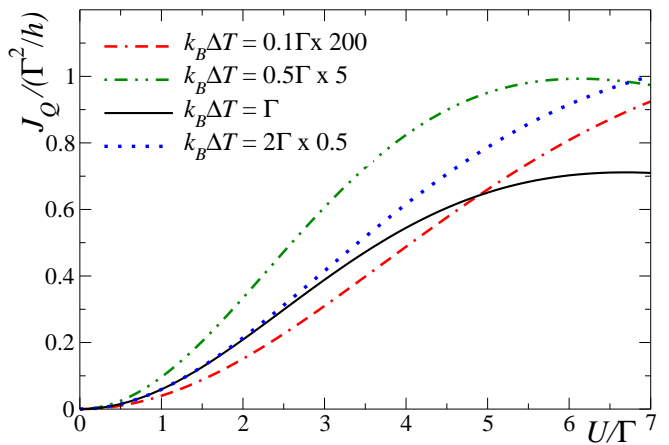}
\caption{(Color online) Thermal current as a function of $U$ for different $\Delta T=T_L$,
$T_R=0$, and $E_L=E_R=-U/2$. Results have been multiplied by a factor in some case, in order to present them in the same scale.}
\label{figtv}
\end{center}
\end{figure}

In Fig. \ref{figtv} we show the thermal current as a function of $U$ calculated with PT in the symmetric configuration $E_L=E_R=-U/2$, for different $\Delta T=T_L$,
keeping $T_R=0$. Since the thermal current strongly depends on $\Delta T$ for 
small $\Delta T$,
the values have been multiplied by a factor indicated in the figure in order to represent them.
In spite of the different magnitude, the different curves show a similar dependence, with 
a $U^2$ behavior for small $U$.  We restrict the range of values  $U$ to those satisfying the criterion of validity of the perturbative approach.
At intermediate $T_L$, 
($0.5 \Gamma$ and $\Gamma$), the curves show a maximum within the interval of $U$ shown.

\begin{figure}[h!]
\begin{center}
\includegraphics[clip,width=\columnwidth]{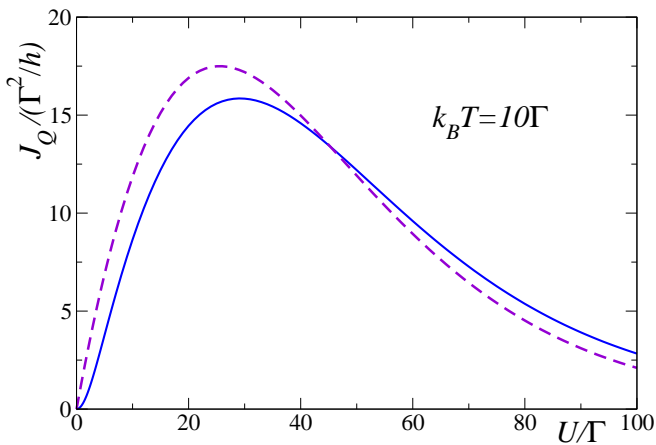}
\caption{(Color online) Same as Fig. \ref{figtv} for $\Delta T=T_L=10 \Gamma$. 
Dashed line corresponds to Eq.  (\ref{jqyada}).}
\label{figt10}
\end{center}
\end{figure}

According to the limit of small $\Gamma$ [Eqs. (\ref{jqf}), (\ref{jqyada})], one expects that for large $\Delta T$ there is a maximum in the thermal current at an intermediate value of $U$. Since at high temperatures, the effects of correlations are expected to be less important, we have also calculated $J_Q$ for $\Delta T=10 \Gamma$ as a function of $U$ for an interval, which includes large values of $U$ lying (at least in principle) beyond the validity of the approach, 
and compare it with the result in the atomic limit 
$\Gamma_\nu \rightarrow 0$ [Eq.  (\ref{jqyada})]. 
The result is shown in Fig. \ref{figt10}. Taking into account the
limitations of both approximations, the results are surprisingly similar. In particular both approaches lead to a maximum in the thermal current for $U \sim 3 \Delta T$. For small $U$ the perturbative approach gives a quadratic dependence in $U$. It is also quadratic in  the atomic limit [Eq.  (\ref{jqf})] for the symmetric case $E_\nu=-U/2$ if in addition both 
$T_L, T_R >0$ but it is linear in $U$ for other cases (Fig. \ref{figt10} 
corresponds to the symmetric case with $T_R=0$).

It is clear that in the atomic limit [Eq.  (\ref{jqf}) or Eq. (\ref{jqyada}) for the symmetric case $E_\nu=-U/2$ plotted in dashed line in Fig. \ref{figt10}], the heat current $J_Q$ vanishes for infinite Coulomb repulsion $U \rightarrow \infty$. While the perturbative result (full line in Fig. \ref{figt10}) lies above the prediction of the analytic expression in the atomic limit for large $U$, perturbation theory
loses its validity for large $U$ and cannot solve the issue of whether $J_Q$ is finite for $U \rightarrow \infty$. However, the NCA result presented in the previous section indicate that
in the symmetric case (keeping $E=-U/2$), $J_Q \rightarrow 0$ for $U \rightarrow \infty$, as discussed previously.

Concerning negative values of $U$, in the symmetric case $E_{\nu}=-U/2$, using the transformation Eq. (\ref{shiba}) one concludes that $J_Q(-U)=J_Q(U)$.
It is easy to check that Eq. (\ref{jqyada}) has this property. We have verified that this is also the case for the perturbative results.

\section{Rectification}
\label{recti}

In the calculations presented before we have considered $E_L=E_R$, although the analytical
results in the atomic limit $\Gamma_\nu \rightarrow 0$ [Eq. (\ref{jqf})] are valid for 
arbitrary $E_\nu$. One effect of having different $E_\nu$ is the loss of the Kondo effect,
in a similar way as the application of a magnetic field in the simplest impurity Anderson model.
Another effect is that the current has a different magnitude when changing the sign of  
$\Delta T$. This rectification effect might be important for applications.\cite{craven}
To keep our convention $T_L > T_R$, we analyze the effect of reflecting the device through the plane that separates the left and right parts, instead of inverting the temperature. 
The result for the magnitude of the current is the same.
Note that if the system has reflection symmetry ($E_L=E_R$, $\Gamma_L=\Gamma_R$, $\mu_L=\mu_R$), 
the magnitude of the current should be unchanged if $\Delta T$ is inverted. Eq. (\ref{jqf}) 
satisfies this symmetry requirement. 

In this Section we keep $\mu_L=\mu_R=0$. Importantly, if the asymmetry is introduced only in the couplings  ($\Gamma_L \neq \Gamma_R$), there is no rectification in the atomic limit, since 
both $\Gamma_\nu$ enter only the prefactor of Eq. (\ref{jqf}). Instead, calculations with the NCA show some rectification effect taking asymmetric couplings but the effect is small and is not reported here.
Concerning PT, for small
$U$ the rectification properties are too small, while for large $U$ the error in the conservation
of the current increased rapidly and we consider that the results were not reliable enough.
Therefore in what follows we also take also $\Gamma_L=\Gamma_R$ and study the effect of different $E_\nu$ in the rectification, using Eq. (\ref{jqf}) for $\Gamma_\nu \rightarrow 0$ and the NCA 
for other cases.

\begin{figure}[h!]
\begin{center}
\includegraphics[clip,width=\columnwidth]{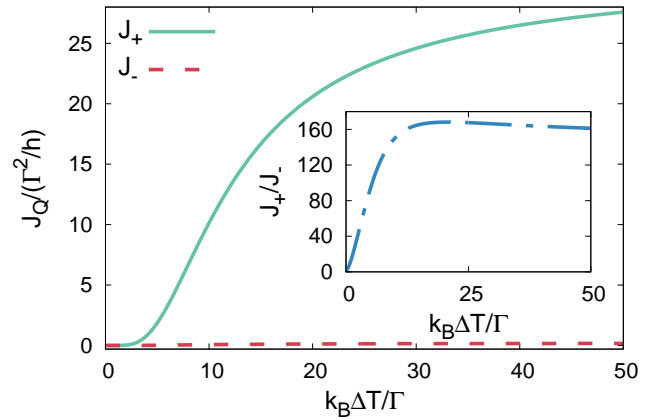}
\caption{(Color online) Thermal currents given by Eq. (\ref{jqf}) as a function of 
$\Delta T$ for $T_R=\Gamma$, $\Gamma_L=\Gamma_R=\Gamma $, $U=20 \Gamma$
and full line $E_L=-14.3 \Gamma$, $E_R=+5\Gamma$ (level nearest to the Fermi energy next to the cold lead), dashed line $E_L=+5\Gamma$, $E_R=-14.3\Gamma$ (level nearest to the Fermi energy next to the hot lead).}
\label{rectif-1}
\end{center}
\end{figure}

In Fig. \ref{rectif-1} we show an example of this rectification effect in the atomic limit.
We have taken one level below and the other one above the Fermi energy. 
We obtain that the magnitude of the heat current is larger, labeled as $J_{+}$ in the figure, 
when the level above the Fermi energy is next to the lead with the lower temperature. 
The opposite direction is labeled as $J_{-}$.
For this choice of parameters, the ratio between both 
currents increases monotonically with $\Delta T$ until $\Delta T\sim 10\Gamma$ and then seems to saturate in a high value of the order of $160$, see inset of figure \ref{rectif-1}. We must warn, however, that such large values of the ratio $J_+/J_-$ are related to the exponentially small  occupancies in the limit $\Gamma_\nu \rightarrow 0$. This effect
disappears for large $\Gamma_\nu$.

\begin{figure}[h!]
\begin{center}
\includegraphics[clip,width=\columnwidth]{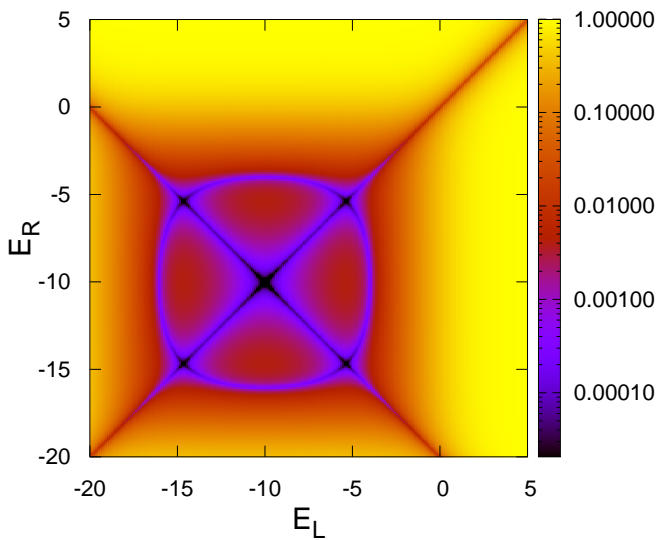}
\caption{(Color online) Rectification coefficient given by Eq. (\ref{r})
calculated in the atomic limit
as a function of both energy levels and $\Delta T=20\Gamma$ and $T_R=\Gamma$ and $U=20 \Gamma$.}
\label{rectif-2}
\end{center}
\end{figure}

The rectification can be quantified by the ratio 
\begin{equation}\label{r}
\mathcal{R}=\big\vert \frac{J_{+}-J_{-}}{J_{+}+J_{-}}\big\vert
\end{equation}
 being $\mathcal{R}=1$ the upper bound. 
In Fig. \ref{rectif-2} we show the values of $\mathcal{R}$ within the atomic limit as a function of both, $E_L$ and $E_R$ for a selected value of $\Delta T=20\Gamma$ while keeping the other parameters as in figure \ref{rectif-1}. Note that the choice of energy levels in Fig. \ref{rectif-1} corresponds to a region in which $\mathcal{R}\sim 1$. 

There are two straight lines in Fig. \ref{rectif-2} that correspond to zero rectification.
In one of them $E_R=E_L$, in which both levels are degenerate, the ratio $\mathcal{R}$ vanishes 
due to reflection symmetry, $L \longleftrightarrow R$.  
The other line $E_R=-U-E_L$ results as a combination of reflection symmetry and the transformation Eq. (\ref{shiba}). In fact the equation for the transformed parameters 
$E_R^\prime=E_L^\prime$, using Eq. (\ref{shibap}) reduces to $E_R=-U-E_L$. 

There is another line of nearly circular shape, in which $\mathcal{R}$ vanishes
which is not related with symmetry properties, 
inside a region of small rectification (and therefore of marginal interest). 
The region inside this line shrinks for increasing 
temperature of the cold lead ($T_R$ in our case). Keeping the other parameters 
of Fig. \ref{rectif-2} fixed, we find that this region  
collapses to the point $E_L=E_R=-U/2$ for $T_R \sim 1.5 \Gamma$.

Near the upper right corner of Fig. \ref{rectif-2} the magnitude of the heat current is larger,
when the level above the Fermi energy is next to the lead with the lower temperature, 
and there is a sign change in $J_+ - J_-$ when crossing the three lines mentioned above.

\begin{figure}[h!]
\begin{center}
\includegraphics[clip,width=\columnwidth]{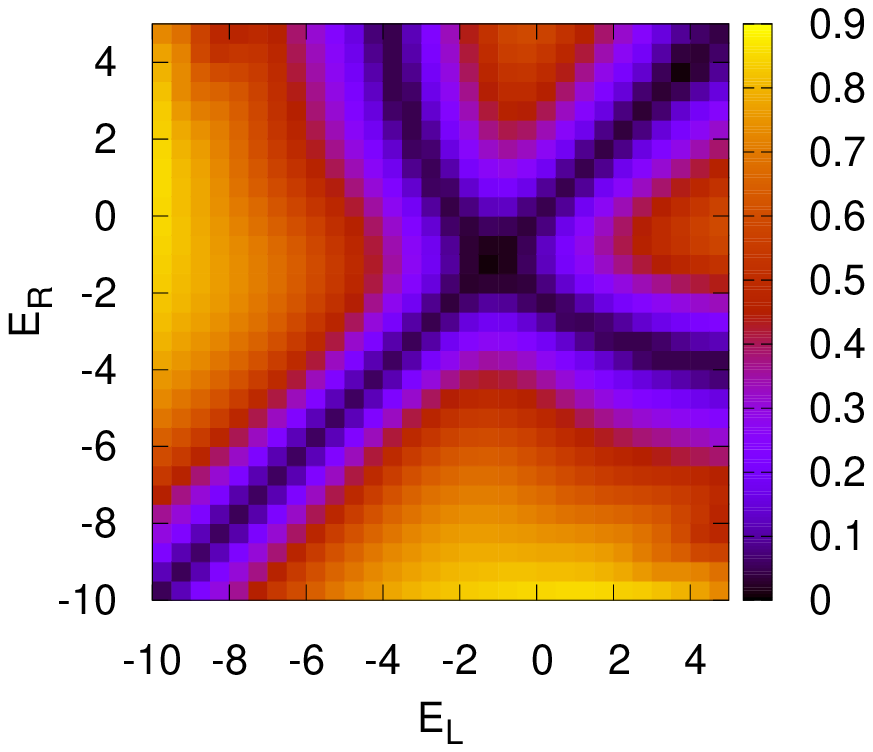}
\includegraphics[clip,width=\columnwidth]{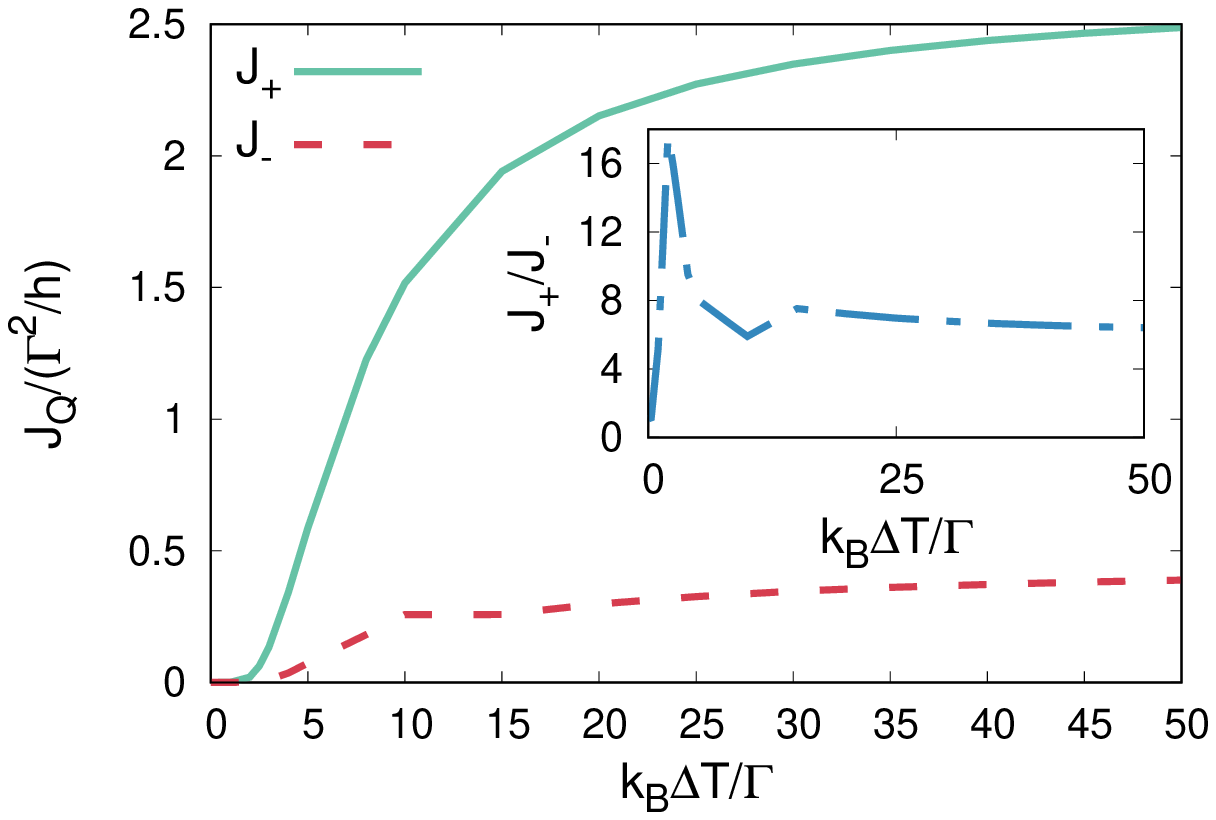}
\caption{(Color online) Top panel: rectification coefficient calculated with NCA 
in the $U \rightarrow \infty$ limit
as a function of both energy levels and $\Delta T=2.5\Gamma$ and $T_R=\Gamma$. Bottom panel: 
Thermal currents as a function of $\Delta T$ for $T_R=0$, $\Gamma_L=\Gamma_R=\Gamma $, $U=\infty$
and full line $E_L=-10 \Gamma$, $E_R=+1.5\Gamma$ (level nearest to the Fermi energy next to the cold lead), dashed line $E_L=+1.5\Gamma$, $E_R=-10\Gamma$ (level nearest to the Fermi energy next to the hot lead). }
\label{recnca}
\end{center}
\end{figure}

In Fig. \ref{recnca} we show the results for $\mathcal{R}$ obtained with the NCA for infinite $U$. 
For regions of parameters where the rectification is important,  
the largest magnitude of the thermal current is obtained when the level nearest to the Fermi
energy is next to the cold lead. This is consistent with the results for the atomic limit presented above.
The behavior of $\mathcal{R}$ as a function of both energy levels is quite similar to the one found within the atomic limit.
In the top panel of Fig. \ref{recnca}, the line $E_L=E_R$ with $\mathcal{R}=0$ is clearly visible and a piece of a curved line with zero rectification also appears. Due to the infinite value of the Coulomb repulsion, the line $E_R=-U-E_L$ with $\mathcal{R}=0$, is not accessible.

Furthermore, the individual currents in both directions, $J_{+}$ and $J_{-}$, display a similar dependence with $\Delta T$ as in the previous case. See the bottom panel of Fig. \ref{recnca}. 
However, the maximum ratio $J_+/J_-$ is reduced from 160 to 16. This is due to that fact that 
the exponentially small  occupancies of some states that take place for $\Gamma_\nu \rightarrow 0$ 
are lost for finite $\Gamma_\nu$.

\section{Summary and discussion}

\label{sum}

We have studied the thermal current through a system of two capacitively coupled quantum dots
connected in series with two conducting leads in the spinless case (corresponding to a high 
applied magnetic field). The system is also equivalent to a molecular quantum dot 
with two relevant levels connected to the leads in such a way that there is perfect destructive interference in the spinless case, and to one spinfull dot between two 
conducting leads fully spin polarized in opposite directions. We expect that our main qualitative results are valid when the spin is included in the former two cases.

An interesting feature of the system is that charge transport is not possible, but heat transport is, due to the effect of the Coulomb repulsion between the electrons in the dots, 
leading to a strong violation of the Wiedemann-Franz law. A simple picture of the effect of 
the Coulomb repulsion in the heat transport is provided in Section \ref{pict}.

The system has been studied previously in the regime of high temperatures of both leads \cite{ruoko,yada}
(including also the full counting statistics \cite{yada}). We extend  those results 
in the limit of small coupling to the leads for arbitrary values of the other parameters. We analyze exhaustively the different regimes of this system,  considering all temperatures and couplings between dots and reservoirs.
In particular, the Kondo regime in which there is one particle strongly localized in the double dot, but fluctuating between both dots. 
For high temperatures of the leads, 
our results agree in general with the previous ones, confirming that the heat current displays
a non-monotonic behavior as a function of Coulomb repulsion and/or coupling to the leads, with a maximum at intermediate values.  

For temperatures $T$ well below the Kondo energy scale $T_K$, we obtain that the thermal
conductance is proportional $T^4$ and the heat current is proportional to $\Delta T^4$,
where $\Delta T$ is the difference between the temperatures of both reservoirs. 
In both cases the behavior changes to linear for $T, \Delta T > T_K$. This implies an important enhancement of the thermal response at low temperatures,
in relation to the case where the coupling between the quantum dots and the reservoirs is very small, where the thermal response is exponentially small.
This property is relevant for the implementation of energy harvesting mechanisms at low temperatures. 

As a function of Coulomb repulsion $U$, for high $\Delta T$ and small tempearature of the cold lead, the heat current has a maximum for $U \sim 3 \Delta T$ and decreases with increasing $U$.
For infinite $U$, we find that the heat current is finite for all non-zero values of 
$\Delta T$ and finite values of the energy levels of the dots $E_\nu$.
Within the Kondo regime, this result can be understood in the frame of renormalized perturbation theory: near the Fermi energy, the main aspects of the physics can be described in terms of dressed weakly interacting quasiparticles. Even if the bare Coulomb repulsion  $U \rightarrow \infty$, 
the renormalized one $\widetilde{U}$ is small and comparable with the renormalized
coupling to the leads. Nevertheless, even at temperatures several orders of magnitude larger
than $T_K$, for which the Kondo effect is destroyed, we obtain a non-zero heat current
for infinite Coulomb repulsion, if $E_\nu$ remains finite.
Instead, in the symmetric case $E_\nu=-U/2$, the current vanishes for $U \rightarrow \infty$.

When the energy levels $E_\nu$ or the the coupling to the leads $\Gamma_\nu$ are different,
the system loses its reflection parity through the plane containing the mid point between  the dots, and therefore, one expects that the absolute value of the heat current $J_Q$ is different for positive or negative temperature difference $\Delta T$. This means that the device has some rectifying properties. In the case in which
only the thermal gradient breaks inversion symmetry one has $J_Q(-\Delta T)=-J_Q(\Delta T)$, 
our results suggest that the asymmetry in the couplings $\Gamma_L \neq \Gamma_R$ modifies the 
amplitude of the current but has little effect on the rectifying properties. Instead, when 
$E_L \neq E_R$, a factor larger than ten between the current flowing in opposite senses can be 
obtained. Our results indicate that the rectification is largest when one level is above and 
near the Fermi energy and the other below the Fermi energy. The largest magnitude of the thermal current is obtained when the former is next to the cold lead.
It is possible that this effect might be increased adding more dots in series.

\section*{Acknowledgments}

We thank Rafael S\'anchez and H. K. Yadalam for helpful discussions.
We are supported by PIP 112-201501-00506 of CONICET and PICT 2013-1045, PICT-2017-2726
of the ANPCyT. LA  also acknowledges  support from
PIP- RD 20141216-4905 of CONICET, CNR-CONICET, and PICT-2014-2049  from Argentina, as well as the Alexander von Humboldt Foundation, Germany.


\begin{thebibliography}{99}
\bibitem{note} An example of ttally destructive interference is benzene molecule doped with one electron or one hole. The many-body ground state
has spin and orbital degeneracy. If the molecule is connected to one lead at an atom 
and the other lead is connected perpendicular to the first one so that it is coupled equally to 
a first and second nearest neighbor of the first atom, there is a linear combination of the
many-body states that couples only to one lead and the other combination to the opposite lead, blocking the transport through the molecule.
The mapping is explained in detail in Ref. \onlinecite{desint} and applies also for other 
molecular quantum dots with destructive interference.
\end{thebibliography}
\end{document}